\documentclass[10pt, conference]{IEEEtran}
\IEEEoverridecommandlockouts
\usepackage{cite}
\usepackage{amsmath,amssymb,amsfonts}
\usepackage{algorithmic}
\usepackage{graphicx}
\usepackage{textcomp}
\usepackage{xcolor}
\def\BibTeX{{\rm B\kern-.05em{\sc i\kern-.025em b}\kern-.08em
    T\kern-.1667em\lower.7ex\hbox{E}\kern-.125emX}}
\begin{document}

\title{Optimizing Stateful Microservice Migration in Kubernetes with MS2M and Forensic Checkpointing 
}

\author{\IEEEauthorblockN{Hai Dinh-Tuan}
\IEEEauthorblockA{Service-Centric Networking \\
\textit{Technische Universit\"at Berlin}\\
Berlin, Germany \\
hai.dinhtuan@tu-berlin.de}
\and
\IEEEauthorblockN{Jialun Jiang}
\IEEEauthorblockA{
\textit{Technische Universit\"at Berlin}\\
Berlin, Germany \\
jialun.jiang@campus.tu-berlin.de
}
}

\maketitle

\begin{abstract}
The widespread adoption of microservices architecture in modern software systems has emphasized the need for efficient management of distributed services. While stateless microservices enable straightforward migration, stateful microservices introduce added complexity due to the need to preserve in-memory state during migration. However, most container orchestrators, including Kubernetes, lack native support for live stateful service migration. This paper proposes an optimized migration scheme for stateful services in Kubernetes by integrating the Message-based Stateful Microservice Migration (MS2M) framework with Kubernetes' Forensic Container Checkpointing (FCC) feature. Key enhancements include support for migrating StatefulSet-managed Pods and the introduction of a \textit{Threshold-Based Cutoff Mechanism} to handle high incoming message rates. Evaluation results demonstrate that MS2M for individual Pods reduces downtime by 96.986\% compared to cold migration methods, while the StatefulSet approach provides greater flexibility in managing stateful services. These insights provide practical strategies for optimizing stateful microservice migration in cloud-native environments.

\end{abstract}


\section{Introduction}

Microservices architecture, a widely adopted architectural style in modern software systems, is based on the philosophy of decomposing monolithic applications into smaller, independently deployable units. Each microservice typically handles a specific function, enabling faster deployments, improved scalability, and more efficient resource utilization. Containers, with their lightweight and isolated environments, have become the preferred deployment method for microservices, often outperforming traditional virtual machines \cite{Singh2017Container-based}. This shift has fueled the widespread adoption of container orchestration platforms, such as Kubernetes, which are now essential for managing the deployment, scaling, and load balancing of microservices \cite{dinh2019maia}.

While the microservices approach offers many advantages, it also introduces significant challenges, particularly in managing service state. The cloud-native paradigm often emphasizes stateless services to simplify management and improve scalability. However, this approach has trade-offs, such as increased latency due to frequent interactions with external databases for state management \cite{Kulkarni2021The}. Unlike stateless services, which can be easily restarted elsewhere, stateful microservices must preserve in-memory state during migration. This adds complexity when containers need to be relocated across different hosts, especially in large-scale applications managed by container orchestration platforms like Kubernetes.

To address these challenges, various approaches for live migration of stateful services have been proposed. While these techniques reduce downtime compared to cold migration methods, they still face limitations, especially in their integration with orchestration platforms like Kubernetes or in high incoming message rate scenarios. This paper builds upon existing frameworks and presents key enhancements to optimize stateful microservice migration in Kubernetes environments. Therefore, the main contributions of this work can be summarized as below:

\begin{enumerate}
    \item \textbf{Integration of MS2M with Kubernetes}: While MS2M \cite{9766576} was originally designed for isolated container migration, this work integrates it with Kubernetes, enabling live migration of Kubernetes-managed Pods.
    
    \item \textbf{Performance Optimization of MS2M}: We introduce a \textit{Threshold-Based Cutoff Mechanism} to address the challenges posed by high incoming message rates, as identified in the original MS2M approach. This optimization reduces the migration time under heavy load, ensuring the process completes efficiently even with increased message traffic.

    \item \textbf{Extension of Forensic Container Checkpointing (FCC)}: We extend Kubernetes' experimental FCC feature to support not just individual containers but also entire Pods, including those managed by Kubernetes controllers like StatefulSet.
\end{enumerate}

The remainder of this paper is organized as follows: Section II reviews the related work, highlighting key contributions and their limitations. Section III outlines the design of our integrated migration approaches, followed by Section IV, which details the technical implementation and evaluation setup. This section also includes an extensive discussion of the evaluation results, assessing the performance of the proposed approaches. Finally, Section V concludes the paper and suggests potential directions for future research.

\section{Related work}

The challenge of service migration in cloud environments, and more broadly across the \textit{Compute Continuum} \cite{balouek2019towards}, has been extensively researched \cite{https://doi.org/10.1002/nem.2212, 8340768}. A recent proposal, Ubiquitous Migration Solution (UMS), outlines a three-phase migration process: pre-migration, migration, and post-migration \cite{10437519}. In the pre-migration phase, the source Coordinator requests migration, and the orchestrator creates a replica container to ensure resources. During migration, control messages are blocked, and a temporary Frontman container notifies users of unavailability. The container is then checkpointed, and its state is transferred. Finally, the source container is deleted, traffic redirected, and the Frontman container removed.

While UMS provides a robust migration framework, its reliance on the \textit{stop-and-copy} method results in extended downtimes, making it unsuitable for time-sensitive applications. This creates a dilemma: stateful microservices are designed to minimize latency by maintaining in-memory state, but \textit{stop-and-copy} negates this advantage by causing significant downtime. As a result, stop-and-copy methods, as used in UMS, contradict the primary purpose of stateful microservices—to reduce latency in dynamic environments.

To address these challenges, various live migration techniques have been developed to keep services running while synchronizing their state. In pre-copy migration, memory is incrementally transferred while the service runs, requiring only a brief pause for final synchronization. For instance, Lu and Jiang \cite{lu2023container} proposed MBDPC (Migration Based on Dirty Page Prediction and Compression), which leverages machine learning to predict modified memory pages, reducing data transfer and downtime. In contrast, post-copy migration suspends the service at the source, resumes it at the destination, and retrieves remaining memory on demand. Tazzioli et al. \cite{tazzioli2024stateful} integrated this approach into Kubernetes, achieving a reported 65\% reduction in downtime. Further optimization is demonstrated by Bellavista et al. \cite{bellavista2024kubernetes}, who separate hot and cold states, significantly minimizing downtime.

A \textit{hybrid pre- and post-copy} method combines pre-copy's incremental transfers with post-copy's on-demand fetching to balance downtime and migration time. For example, Ma et al. \cite{ma2018efficient} proposed a live migration method for Docker containers in edge environments, reducing file system synchronization by transferring only the top writable layer, cutting migration time by 56\% to 80\% under varying network conditions.

While the reviewed migration techniques are broadly applicable, they require direct handling of in-memory state, with optimizations relying on careful state management, as demonstrated in \cite{bellavista2024kubernetes}. A simpler and more efficient approach to state management can be derived from the way microservices are designed, many of which adopt an event-driven architecture. In such designs, a microservice's state changes only in response to events or messages \cite{Rahmatulloh2022Event-Driven}, enabling service state to be reconstructed through message exchanges rather than direct manipulation of in-memory state.

Building on this principle, Dinh-Tuan and Beierle \cite{9766576} introduced MS2M, a live migration technique specifically tailored for stateful microservices across the compute continuum. MS2M leverages the existing message exchange infrastructure to synchronize service states during migration. Instead of transferring in-memory state, it reconstructs the service state at the destination by replaying cached incoming messages. To ensure seamless migration, a \textit{secondary message queue} temporarily stores incoming messages, replaying them to maintain state synchronization with minimal downtime. Experimental results show that MS2M significantly reduces downtime compared to traditional stop-and-copy methods, making it particularly effective for distributed systems that rely on asynchronous messaging.

\section{Concept and Design}

\subsection{Migration Manager and the migration workflow}

\begin{figure}[htbp]
\centering
\includegraphics[width=0.5\textwidth]{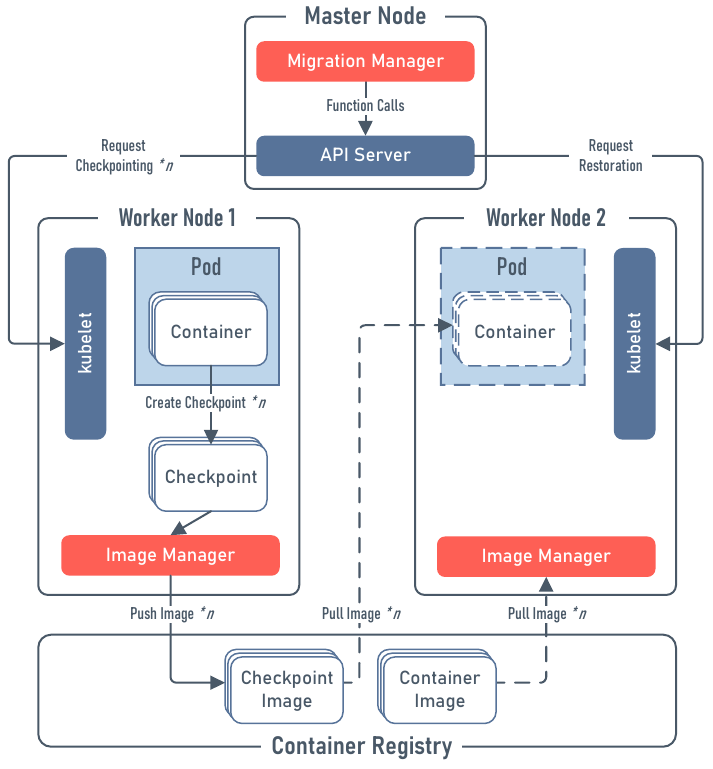}
\caption{Checkpoint/Restore Workflow}
\label{fig: Checkpoint/restore}
\end{figure}

As illustrated in Figure \ref{fig: Checkpoint/restore}, the red-colored components highlight the additional components introduced in our proposal. The \textit{Migration Manager} is a custom-built program that leverages Kubernetes' experimental checkpointing feature, FCC, to automate the migration of stateful applications. Deployed on the Kubernetes master node, it interacts directly with the Kubernetes control plane via the API Server. The Migration Manager integrates the MS2M approach into Kubernetes and extends its framework to support Pod-level migrations, a capability not yet natively supported. Upon receiving a migration request, it retrieves the Pod’s configuration and gathers container information to ensure consistency. It then directs the control plane to create checkpoints for each container, which are compiled into checkpoint images by the \textit{Image Manager} and stored in the container registry. 

\subsection{Migration of individual pods}

The migration process for individual Pods builds upon the original MS2M design \cite{9766576}, with key adaptations for Kubernetes integration. Initially, the process involved three main actors: the Source Host, Target Host, and Migration Manager. To integrate with Kubernetes, the API Server (AS) was added as the fourth actor, redirecting interactions from direct host communication to the Kubernetes Control Plane (KCP), which manages Kubernetes Nodes. This integration preserves Kubernetes' resource management model, allowing it to manage tasks such as scheduling and lifecycle operations seamlessly.

\begin{figure}[htbp]
\centering
\includegraphics[width=0.5\textwidth]{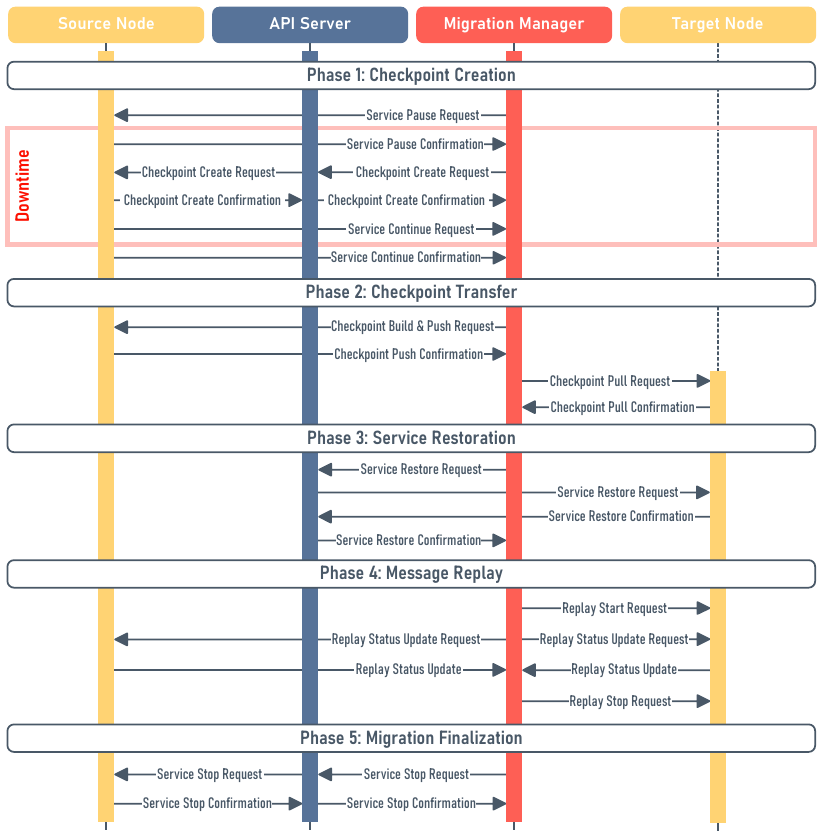}
\caption{Detailed Migration Process of MS2M for individual Pods}
\label{fig: Detailed migration process of MS2M for individual Pods}
\end{figure}

As shown in Figure \ref{fig: Detailed migration process of MS2M for individual Pods}, during migration, services are paused and restored through interactions with the AS, which directs the Kubernetes Nodes hosting the Pods. Checkpointing is performed using Kubernetes' FCC feature. To initiate the process, a request is sent to the AS, which locates the containers within the specified Pod and instructs the Nodes to generate the necessary checkpoints.

Instead of transferring checkpoints directly, the process involves building and pushing checkpoint images from the source Node to a container registry. This decouples the source and target Nodes, optimizing network usage and providing greater flexibility for scheduling the restoration process. Independent of the source Node’s availability, the target Node pulls the images from the registry to restore the service locally when those images are available.

After the newly restored service has finished synchronizing its state, the service on the source Node is permanently stopped by deleting the Pod via a request to the AS. This ensures the service is fully terminated on the source Node, allowing the migration process to proceed without any disruption.

\begin{figure}[htbp]
    \centering
    \includegraphics[width=0.5\textwidth]{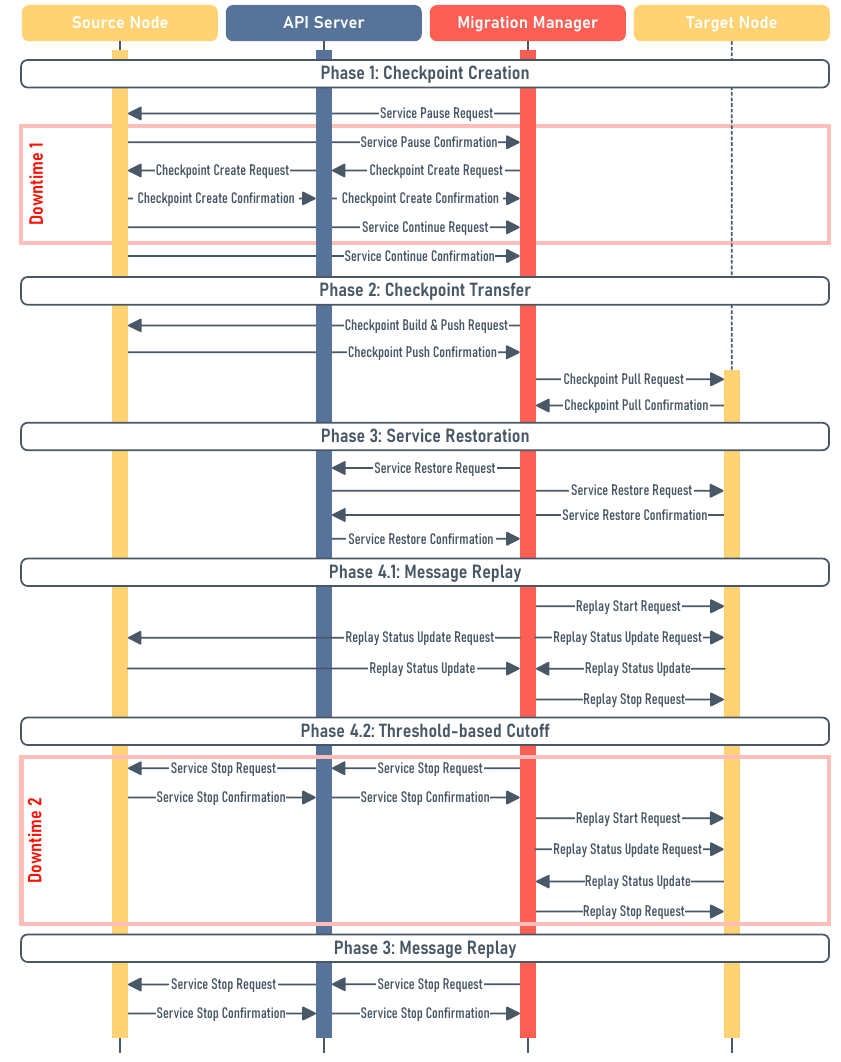}
    \caption{Detailed migration process of MS2M for individual Pods with Threshold-Based Cutoff Mechanism}
    \label{fig: Detailed migration process of MS2M for individual Pods with Threshold-Based Cutoff Mechanism}
\end{figure}

However, the aforementioned process does not fully address the challenges associated with the original MS2M design, particularly in scenarios where the incoming message rate is exceptionally high. In such cases, performance can degrade significantly as the newly migrated Pod struggles to process the large volume of messages accumulating in the secondary queue for state synchronization. In extreme cases, where the incoming message rate nears the Pod's processing capacity, the migration process may even fail to complete successfully \cite{9766576}. To resolve this, we propose a \textit{Threshold-Based Cutoff Mechanism}. This approach involves stopping the source Pod and allowing the newly migrated Pod to replay messages until its state aligns with the final state of the source Pod before deletion. Once the source Pod is stopped, any incoming messages are not processed immediately but are stored in the message queue, to be handled after the migration process is completed.

To directly control the synchronization duration, we formulate the calculation of the \textit{cutoff threshold} ($T_{\text{cutoff}}$) based on the \textit{maximum acceptable message replay time} ($T_{\text{replay\_max}}$), utilizing principles from queuing theory. This approach ensures that the time spent replaying messages at the target microservice remains within acceptable limits, preventing excessive resource consumption and prolonged delays in service availability. Consequently, it provides predictable migration performance and balances the trade-off between service downtime and resource utilization.

From a queuing theory perspective, the microservice processing can be modeled as an M/M/1 system. The key parameters for this model are defined as follows:

\begin{itemize}
    \item $\lambda$: Incoming message rate at the source microservice (messages per second).
    \item $\mu_{\text{target}}$: Message processing rate at the target microservice (messages per second).
    \item $T_{\text{replay\_max}}$: Maximum acceptable time for message replay at the target microservice.
    \item $T_{\text{accum}}$: Time during which messages accumulate in the secondary queue.
\end{itemize}

During migration, messages arrive at the source microservice following a Poisson process with rate $\lambda$ and accumulate in the secondary queue for a duration $T_{\text{accum}}$, resulting in a total number of messages:

\begin{equation}
N_{\text{messages}} = \lambda \times T_{\text{accum}}.
\end{equation}

The time required to replay these messages at the target microservice is:

\begin{equation}
T_{\text{replay}} = \frac{N_{\text{messages}}}{\mu_{\text{target}}} = \frac{\lambda \times T_{\text{accum}}}{\mu_{\text{target}}}.
\end{equation}

To ensure that $T_{\text{replay}}$ does not exceed $T_{\text{replay\_max}}$, we impose the condition:

\begin{equation}
T_{\text{replay}} \leq T_{\text{replay\_max}}.
\end{equation}

Substituting the expression for $T_{\text{replay}}$, we obtain:

\begin{equation}
\frac{\lambda \times T_{\text{accum}}}{\mu_{\text{target}}} \leq T_{\text{replay\_max}}.
\end{equation}

Solving for $T_{\text{accum}}$, we derive the cutoff threshold:

\begin{equation}
T_{\text{cutoff}} = T_{\text{accum}} \leq \frac{T_{\text{replay\_max}} \times \mu_{\text{target}}}{\lambda}.
\end{equation}

This cutoff threshold ensures that the replay time remains within acceptable limits by controlling the accumulation period based on the system's message arrival and processing rates. As shown in Figure~\ref{fig: Detailed migration process of MS2M for individual Pods with Threshold-Based Cutoff Mechanism}, the \textit{Threshold-Based Cutoff Mechanism} is triggered when $T_{\text{cutoff}}$ expires, preventing prolonged or excessive message replay that could otherwise degrade performance or delay service availability.

\subsection{Migration of Kubernetes controlled pods}

Kubernetes operates on the principle of \textit{desired state}, which is managed through objects like \textit{Pods}, \textit{Deployments}, and \textit{StatefulSets}. The \textit{kube-controller-manager daemon} continuously monitors the cluster and ensures that the actual state matches the desired state, taking corrective actions such as recreating failed Pods.

\begin{figure}[htbp]
    \centering
    \includegraphics[width=0.5\textwidth]{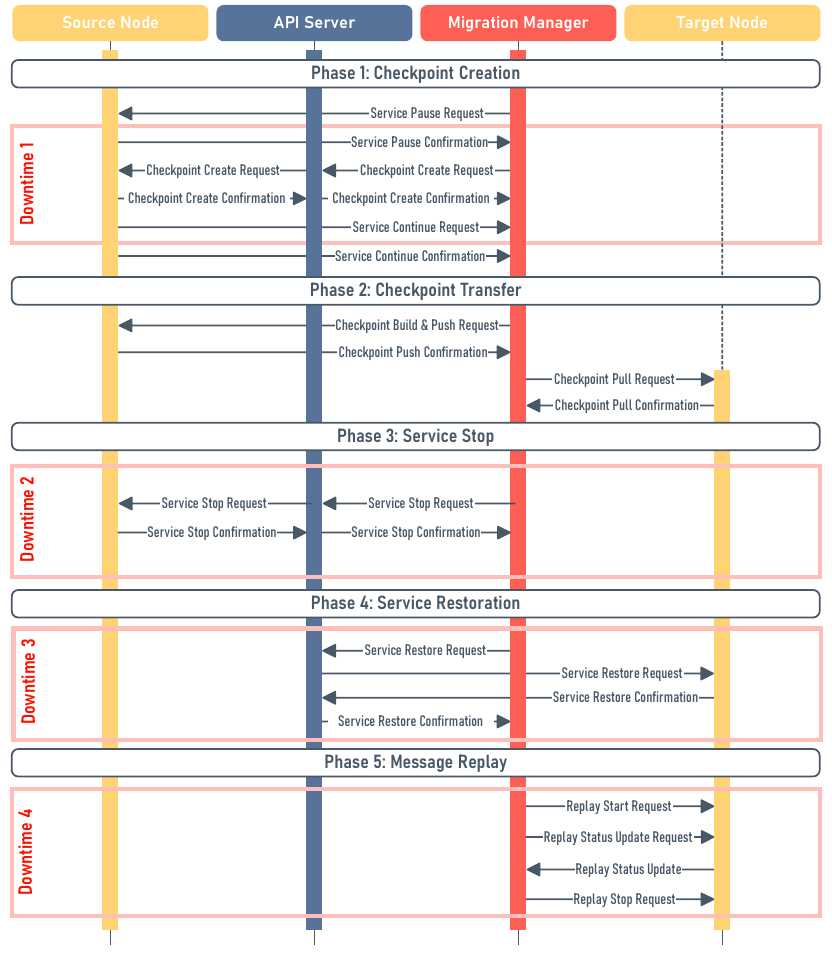}
    \caption{Detailed migration process of MS2M for StatefulSet Pods}
    \label{fig: Detailed migration process of MS2M for StatefulSet Pods}
\end{figure}

StatefulSet is specifically designed for managing stateful applications by providing each Pod with a stable network identity, persistent storage, and ensuring sequential deployment and scaling. These features are critical for stateful services such as databases, which require consistent data and identity across Pods. However, this introduces challenges for migration, as each Pod in a StatefulSet has a permanent, unique identity and address. This makes it impossible to keep the source Pod running while creating a copy for migration. Consequently, MS2M concept cannot be directly applied, as the source Pod must be deleted before the target Pod can be created. This restriction complicates live migration because maintaining unique Pod identities prevents simultaneous operation of both source and target Pods.

In the modified MS2M process supporting StatefulSet (illustrated in Figure \ref{fig: Detailed migration process of MS2M for StatefulSet Pods}), the service is stopped after the checkpoint transfer phase. Only the target Pod is involved in the message replay, processing messages up to the \textit{cutoff message ID}, which corresponds to the last message received by the source Pod before it was stopped. Similar to the cutoff mechanism explained above, while this method reduces migration time, it results in increased downtime, a necessary trade-off for StatefulSet-managed Pods.

It is also worth mentioning that for stateful applications in Kubernetes, data partitioning is often used  to divide data and workload across multiple stateful service instances. For example, a message queue might be partitioned based on certain keys, with each partition assigned to a specific instance of the stateful service. This ensures that each instance processes only the subset of messages relevant to its state. Or in some cases, instead of having all service instances subscribe to the same message queue, stateful services may require dedicated message queues for each instance. For example, each Pod in a StatefulSet may have its own queue, allowing it to process messages related only to its specific state or responsibility.

\section{Implementation and evaluation}

\subsection{Implementation setup}

For our implementation, the microservices are developed in \textit{Java 17} with \textit{Spring Boot 3.1.5}, while the Migration Manager, responsible for automating migration and deployment, is implemented in \textit{Python 3.10}. Two types of microservices were deployed in the cluster:
\begin{itemize}
    \item \textit{Producer}: A single-instance microservice deployed in a Pod, responsible for continuously sending messages to RabbitMQ or based on specific triggers.
    \item \textit{Consumer}: A scalable, multi-instance microservice deployed using StatefulSet, responsible for consuming messages from RabbitMQ queues, ensuring the application can handle varying workloads and scale as needed.
\end{itemize}

The system was deployed on \textit{Google Compute Engine}, comprising three \textit{e2-medium} virtual machines (2 vCPUs, 4 GB memory each) using \textit{Ubuntu 20.04 Focal Fossa} with \textit{cgroup v1} for compatibility with checkpointing. \textit{Kubernetes v1.30} orchestrated the containerized services, using \textit{CRI-O v1.28} as the container runtime, with \textit{CRIU} enabled for checkpoint/restore functionality. To transfer checkpoint files, \textit{Rsync} over \textit{SSH} was used, and \textit{Buildah 1.37.0} created Open Container Initiative (OCI)-compliant images for migration. \textit{Google Cloud’s Artifact Registry} stored both container and checkpoint images, while \textit{Persistent Volumes} ensured data consistency. \textit{RabbitMQ v3.13} was deployed as a message broker to manage inter-service communication.

\subsection{Evaluation Results}

In this evaluation, we compare the performance of four migration approaches: the baseline Stop-and-Copy migration and three MS2M-based strategies. The evaluated strategies are: Stop-and-Copy Migration, MS2M for Individual Pods, MS2M for Individual Pods with the Threshold-Based Cutoff Mechanism, and MS2M for StatefulSet Pods.

We focus on two key parameters to assess performance: the \textit{message rate} (number of messages sent per second by the producer) to measure how each approach handles varying workloads, and the \textit{message processing time} (average time for the consumer to process a message), simulated by introducing delays. A baseline was set with a message rate of 10 messages per second and a processing time of 50 milliseconds (resulting in a maximum processing rate of 20 messages per second). The message rate was then varied to evaluate its impact, while keeping processing time constant. Each migration strategy was tested with different message rates, with each test case run 10 times.

We used two primary metrics to assess performance, in line with similar research \cite{Sun2016A, Wu2017Live, 9766576}:
\begin{itemize}
    \item \textbf{Total Migration Time}: The duration from the service pause request at the source node until the microservice is fully restored and synchronized on the target node.
    \item \textbf{Downtime}: The period during which the consumer microservice is unable to process messages.
\end{itemize}

\begin{figure}[htbp]
    \centering
    \includegraphics[width=0.45\textwidth]{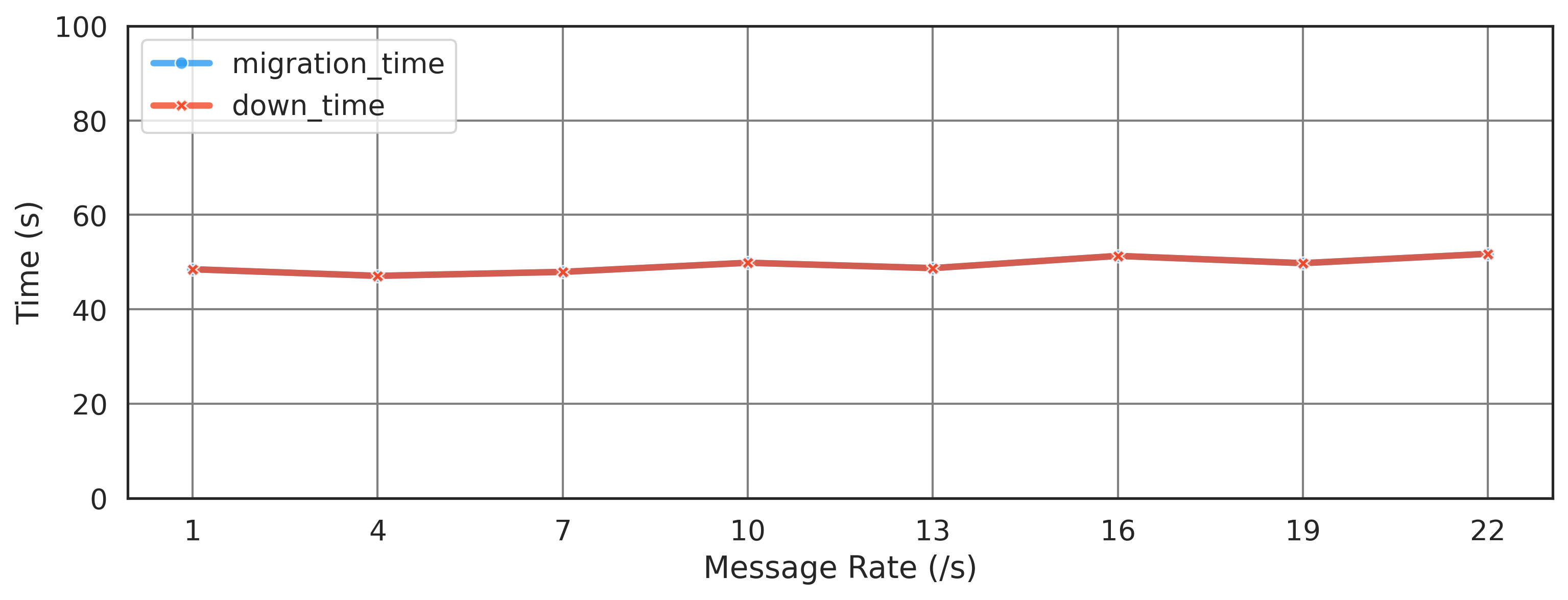}
    \caption{The baseline stop-and-copy migration.}
    \label{fig:cold_migration_avg_time}
\end{figure}

\begin{figure}[htbp]
    \centering
    \includegraphics[width=0.45\textwidth]{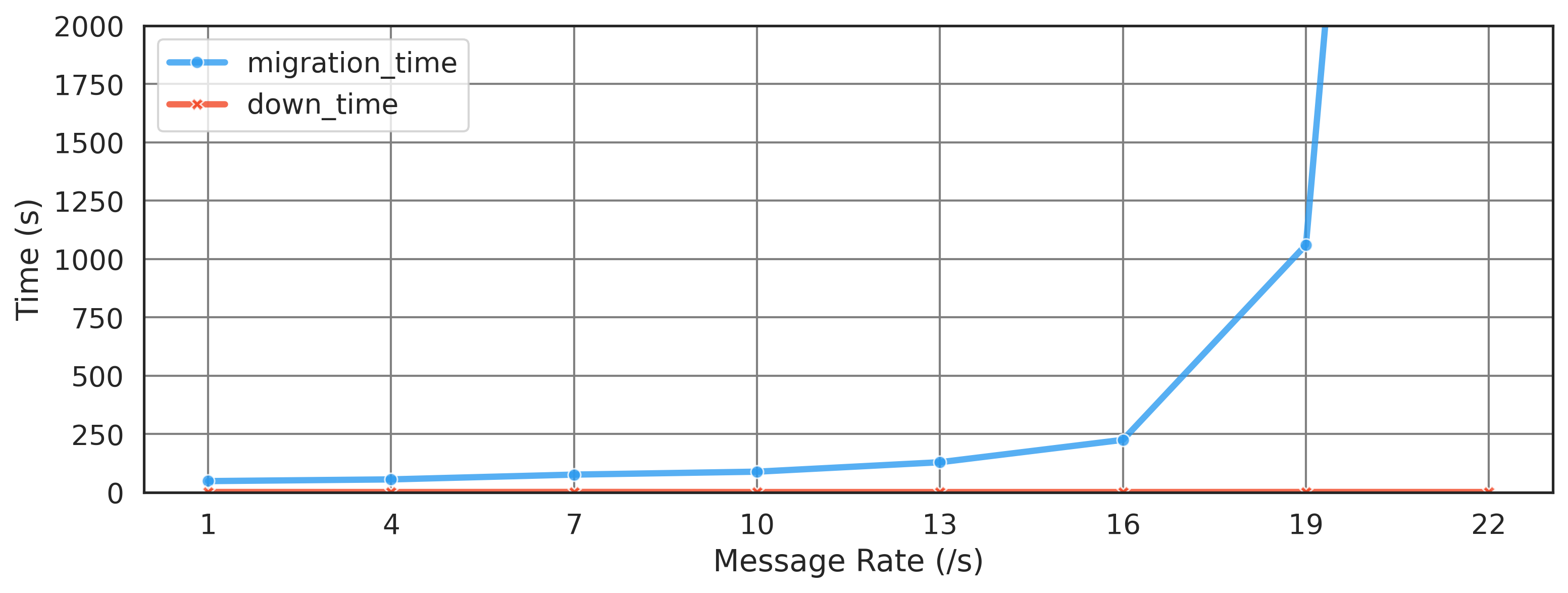}
    \caption{MS2M for Individual Pods.}
    \label{fig:individual_pods_avg_time}
\end{figure}

\begin{figure}[htbp]
    \centering
    \includegraphics[width=0.45\textwidth]{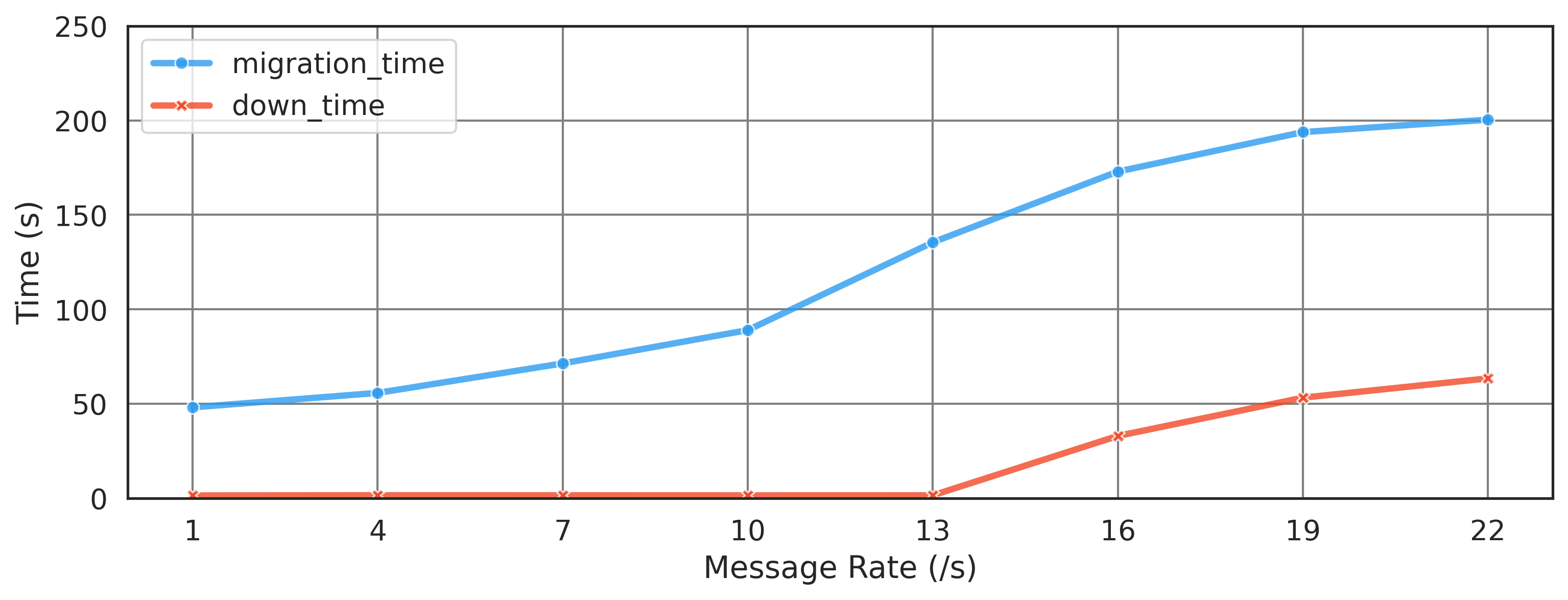}
    \caption{MS2M for Individual Pods with Cutoff mechanism.}
    \label{fig:individual_pods_early_stop_avg_time}
\end{figure}

\begin{figure}[htbp]
    \centering
    \includegraphics[width=0.45\textwidth]{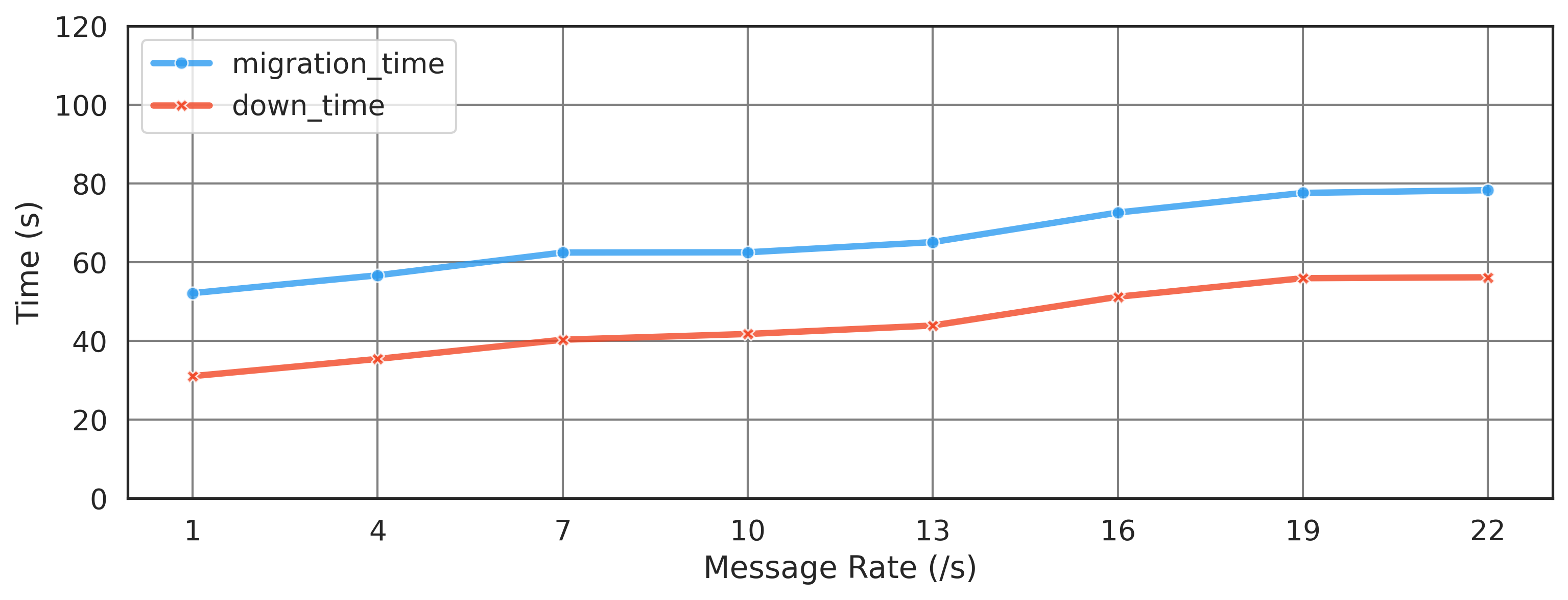}
    \caption{MS2M for StatefulSet Pods.}
    \label{fig:ms2m_stateful_avg_time}
\end{figure}

For stop-and-copy migration, the migration time remains relatively constant across different message rates, averaging 49.055 seconds, as shown in Figure \ref{fig:cold_migration_avg_time}. This is expected since the service is fully suspended throughout the migration process, resulting in downtime that equals the migration time. Since no message-based synchronization method is used, the message rate does not impact the migration time. As previously discussed, this characteristic makes Stop-and-Copy unsuitable for high-availability scenarios, reinforcing the need for live migration techniques like MS2M.

In the case of MS2M for individual Pods (Figure \ref{fig:individual_pods_avg_time}), migration time increases significantly as the message rate approaches the maximum processing rate of 20 messages per second. This indicates that this method struggles to scale effectively under high traffic. However, downtime remains consistently low at an average of 1.547 seconds (a 96.846\% reduction), demonstrating that MS2M effectively minimizes service unavailability despite longer migration times at higher message rates.

For MS2M for individual Pods with Threshold-based Cutoff Mechanism enabled (Figure \ref{fig:individual_pods_early_stop_avg_time}), the migration time increases more gradually as the message rate rises, indicating better handling of higher traffic compared to the standard MS2M approach. While downtime also increases when the cutoff mechanism activates, the rate of increase is slower than that of migration time, suggesting that the cutoff mechanism effectively reduces total migration time by cutting off message replay when necessary. This makes it more efficient under high message rates.

Finally, MS2M for StatefulSet Pods shows a more moderate increase in both migration time and downtime as message rates rise as illustrated in Figure \ref{fig:ms2m_stateful_avg_time}. While both metrics increase, the rise is more controlled compared to individual Pods, suggesting that MS2M for StatefulSet Pods handles higher message rates more gracefully, although service availability is still impacted as message rates increase.

In summary, Stop-and-Copy migration is the least flexible and unsuitable for high-availability scenarios. MS2M for Individual Pods excels in keeping downtime low but struggles with long migration times at higher message rates. MS2M with the cutoff mechanism offers a better balance, reducing migration time spikes while maintaining acceptable downtime. MS2M for StatefulSet Pods performs well under increasing message rates, handling higher traffic effectively but still experiencing increased downtime and migration time.

\begin{figure}[htbp]
    \centering
    \includegraphics[width=0.45\textwidth]{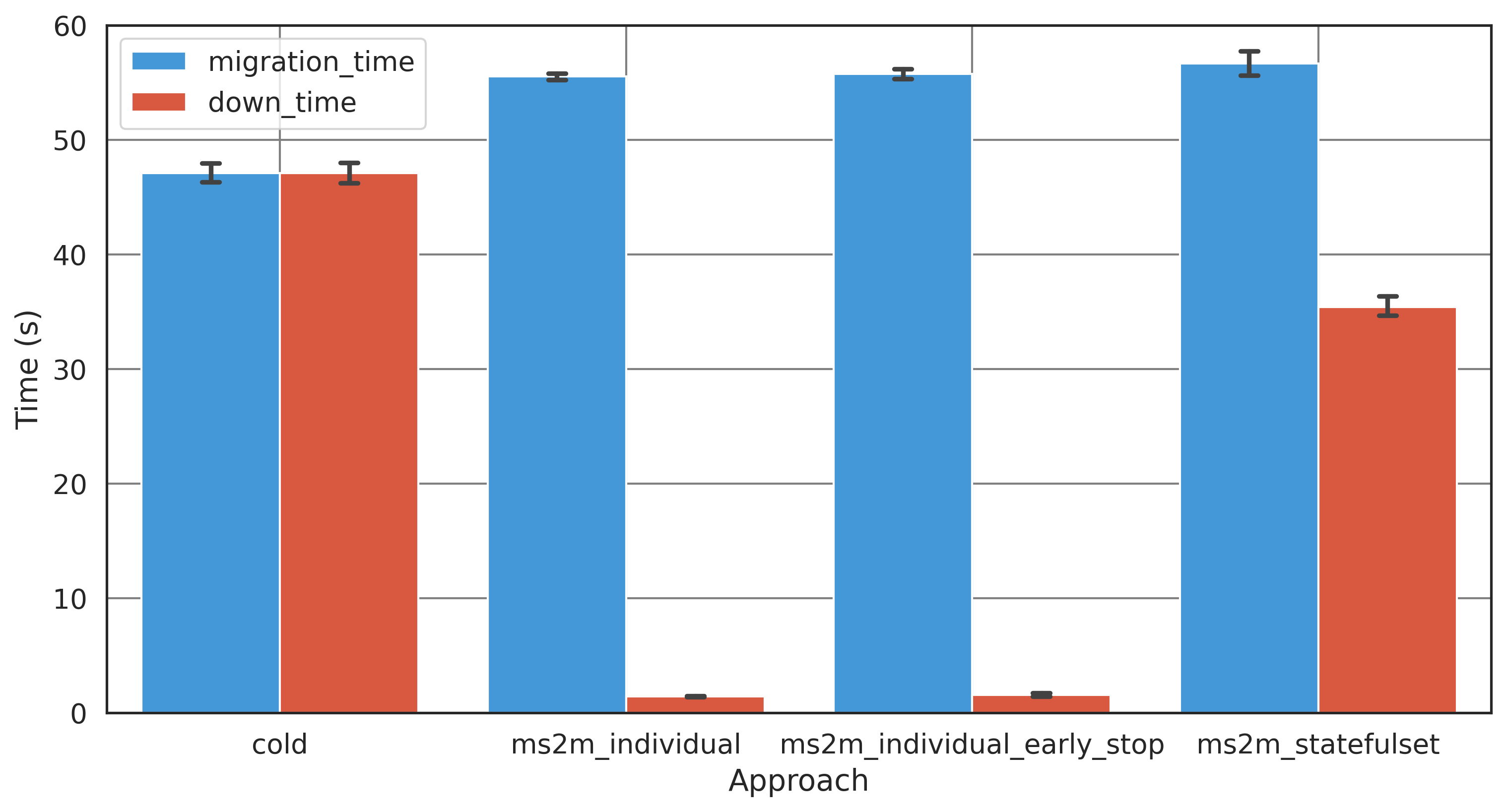}
    \caption{Low Message Rate (4/s)}
    \label{fig:low_message_rate_04_avg_time}
\end{figure}

\begin{figure}[htbp]
    \centering
    \includegraphics[width=0.45\textwidth]{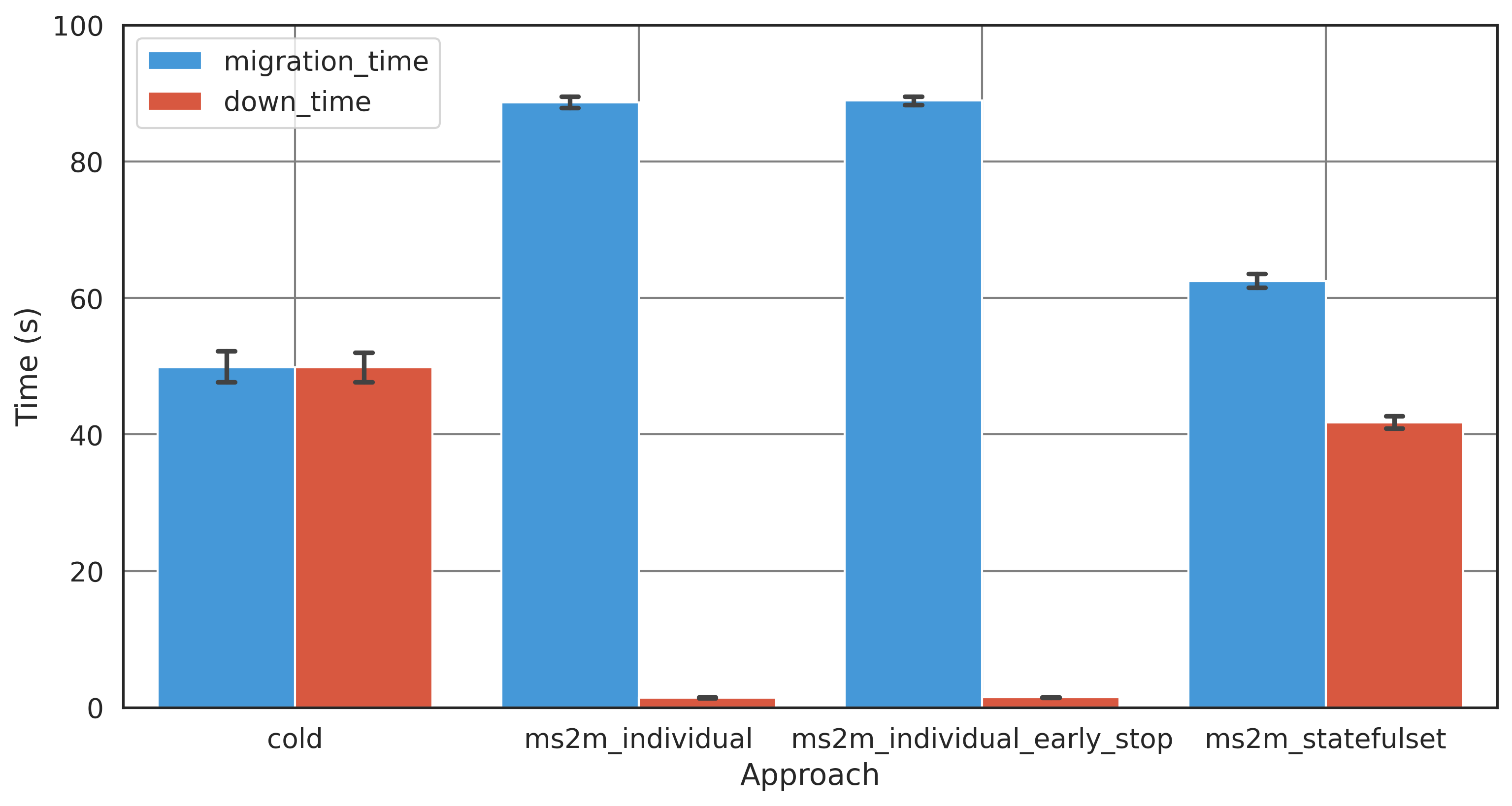}
    \caption{Intermediate Message Rate (10/s)}
    \label{fig:intermediate_message_rate_10_avg_time}
\end{figure}

\begin{figure}[htbp]
    \centering
    \includegraphics[width=0.45\textwidth]{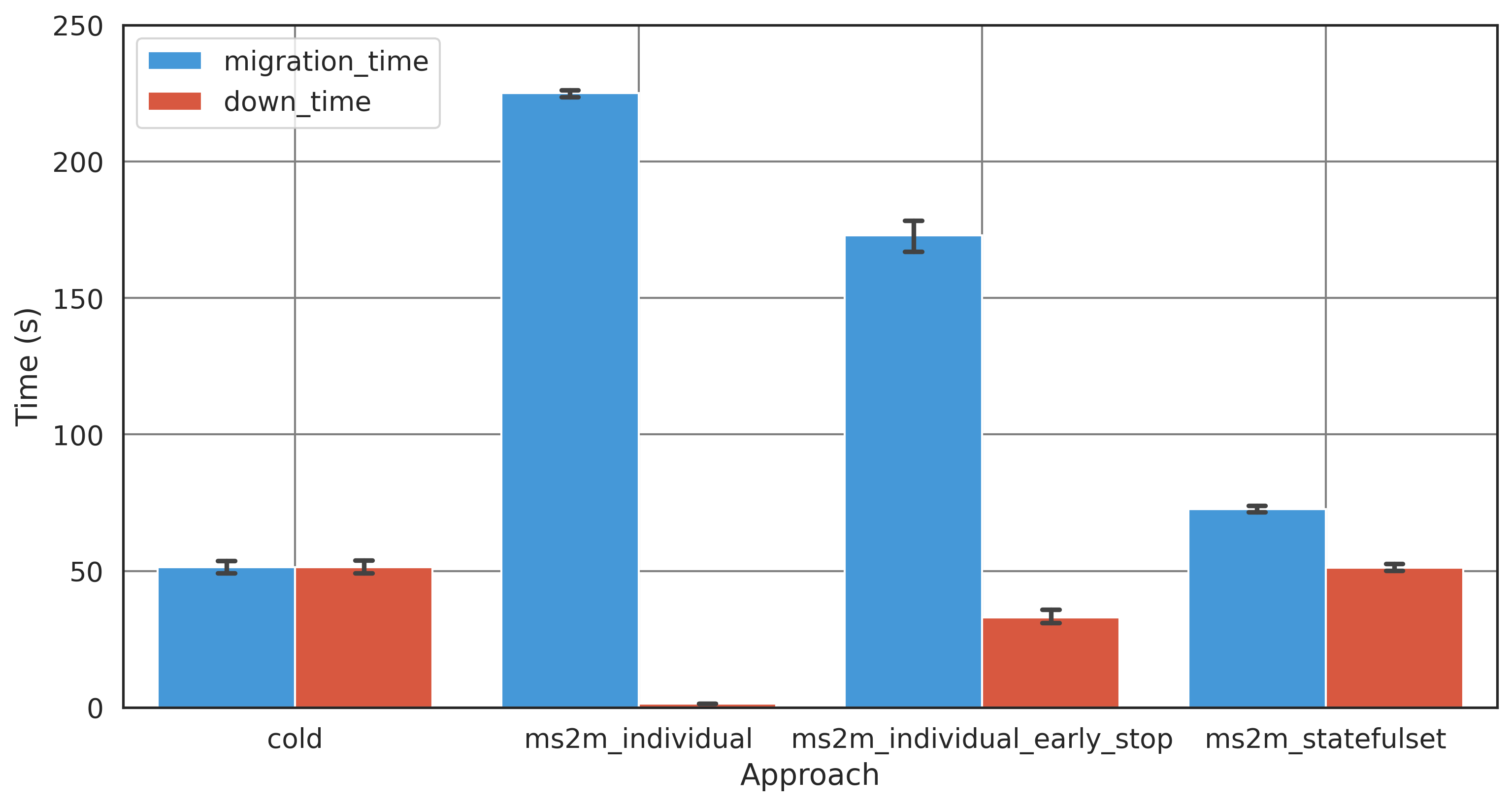}
    \caption{High Message Rate (16/s)}
    \label{fig:high_message_rate_16_avg_time}
\end{figure}

We also compared the performance of each migration strategy across three message rate scenarios, as illustrated in Figures \ref{fig:low_message_rate_04_avg_time}, \ref{fig:intermediate_message_rate_10_avg_time}, and \ref{fig:high_message_rate_16_avg_time}.

At low message rates, the total migration time increases, while downtime decreases most noticeably during the migration of individual pods, particularly when compared to the baseline Stop-and-Copy migration (47.077 seconds). This supports the primary argument that MS2M significantly reduces service unavailability during migration. However, the effects of MS2M are less pronounced for Stateful-managed pods. For individual pods, the downtime is reduced by 96.986\%, while the total migration time increases by 17.981\%. When using the cutoff mechanism, the downtime is reduced by 96.737\%, with a similar increase in total migration time. For StatefulSet pods, downtime decreases by 24.840\%, though total migration time increases by 20.307\%. 

At intermediate message rates, during individual pod migration, downtime remains very low, but migration time increases significantly due to prolonged message replay and state synchronization. Despite this, the downtime remains much shorter than in Stop-and-Copy. For individual pods, the downtime is reduced by 97.178\%, with a substantial increase in total migration time. With the cutoff mechanism, the downtime is reduced by 97.047\%, with a slight further increase in total migration time. Migration of StatefulSet pods also takes longer because of message replay, and as this time grows, the gap in downtime between stop-and-copy migration and StatefulSet migration narrows. In this case, the downtime for StatefulSet pods is reduced by 16.309\%.

At high message rates, the individual pod approach without the cutoff mechanism maintains low downtime, but results in the longest migration time. The cutoff mechanism reduces migration time at the cost of slightly higher downtime. For individual pods, downtime drops by 97.178\%, but migration time increases significantly. With the cutoff mechanism, downtime reduction falls to 36.076\%, though migration time improves. For StatefulSet pods, downtime becomes comparable to stop-and-copy migration, indicating that the benefits of message replay diminish at high message rates. Specifically, in this case, the downtime remains relatively unchanged, decreasing by only 0.242\%, while the total migration time increases by 41.466\%.

Overall, message rates increase, both total migration time and downtime rise across all scenarios. MS2M for individual Pods, with or without the Cutoff Mechanism, significantly reduces downtime compared to stop-and-copy, with the cutoff mechanism offering additional benefis as it balances migration time and downtime. In the StatefulSet scenario, MS2M's effectiveness in reducing downtime lessens as message rates increase, showing that higher rates erode its advantages.

\begin{figure}[htbp]
    \centering
    \includegraphics[width=0.4\textwidth]{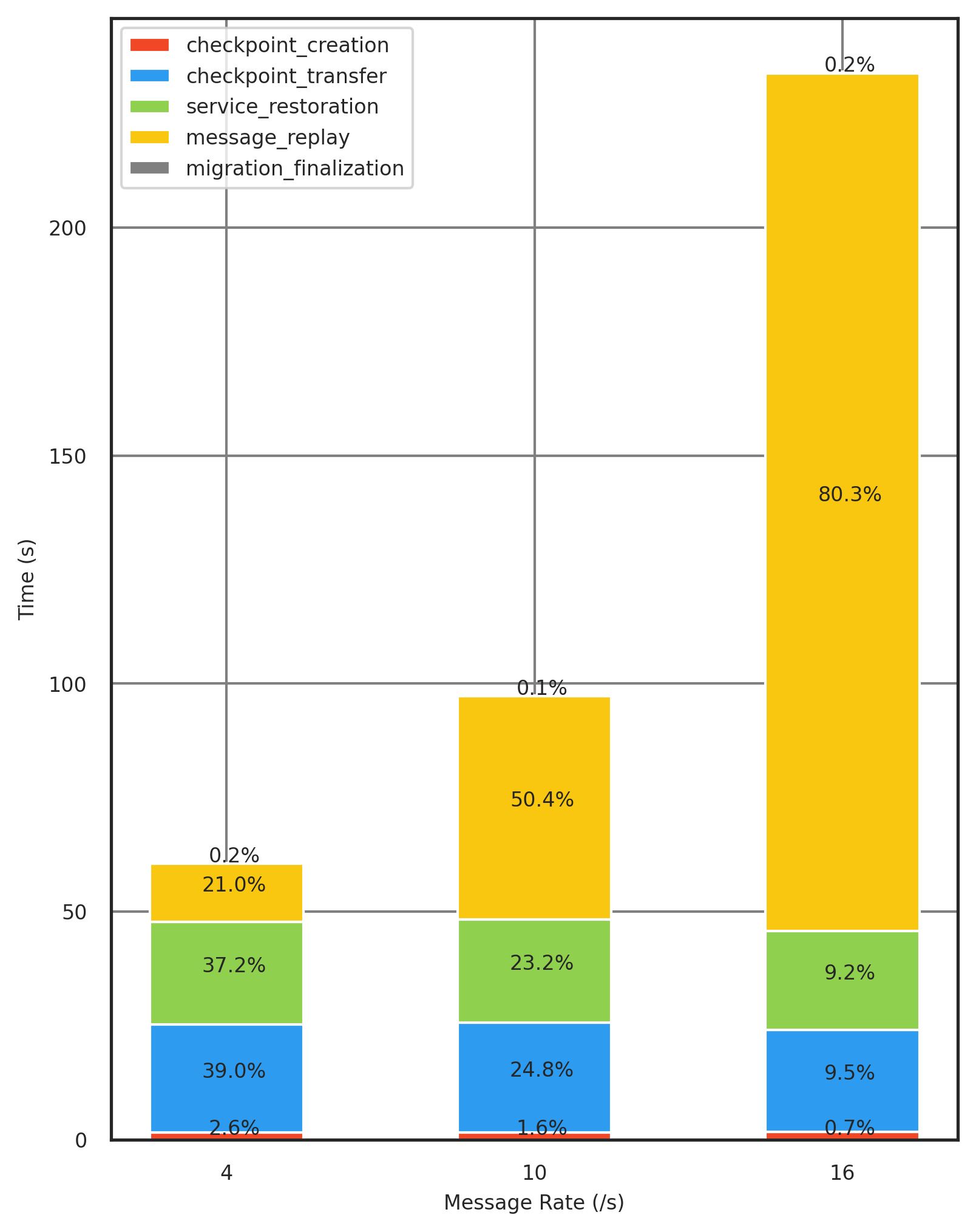}
    \caption{MS2M for Individual Pods}
    \label{fig: Time Distribution Across Migration Phases - MS2M for Individual Pods}
\end{figure}

\begin{figure}[htbp]
    \centering
    \includegraphics[width=0.4\textwidth]{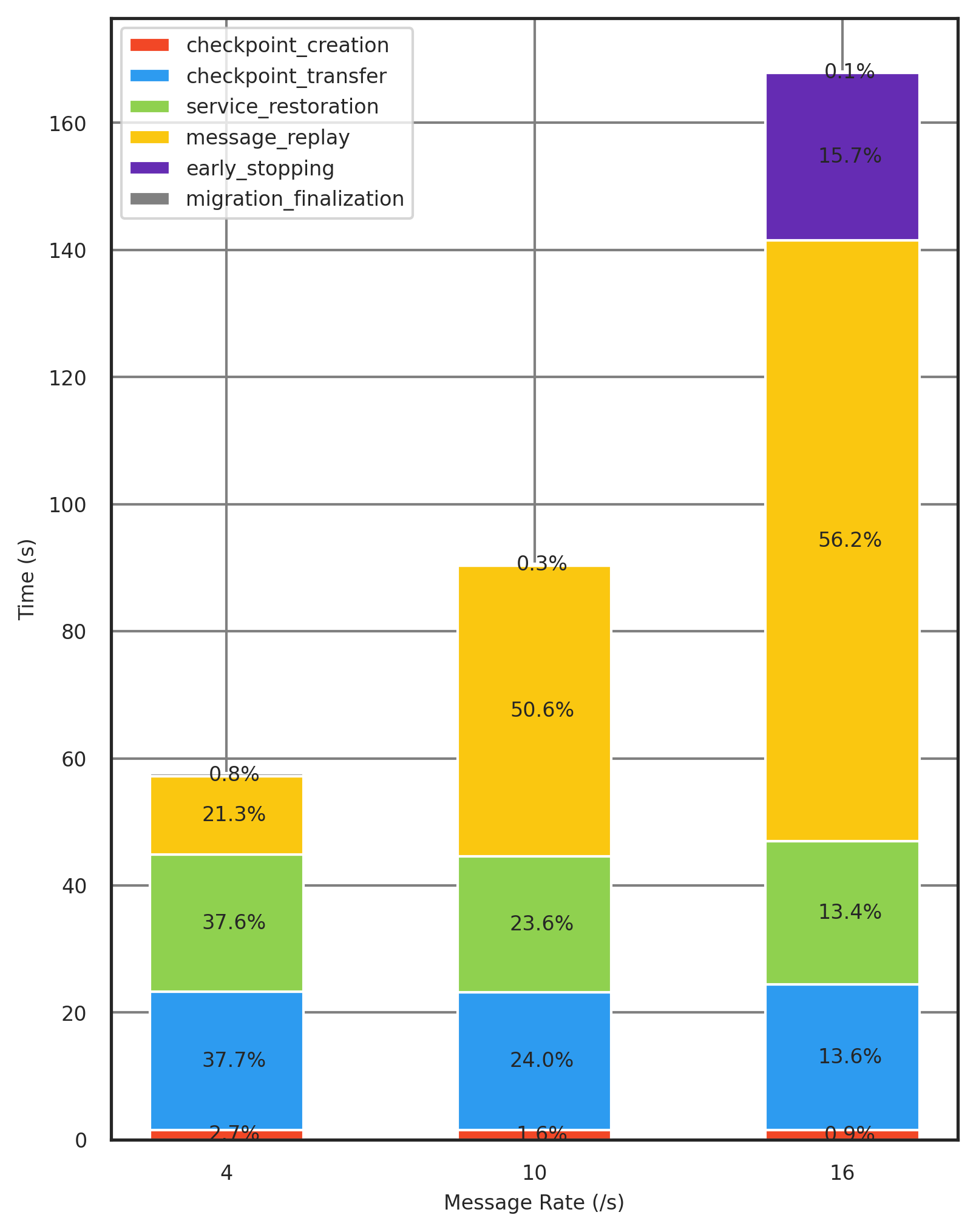}
    \caption{MS2M for Individual Pods with Cutoff Mechanism}
    \label{fig: Time Distribution Across Migration Phases - MS2M for Individual Pods with Early Stopping}
\end{figure}

\begin{figure}[htbp]
    \centering
    \includegraphics[width=0.4\textwidth]{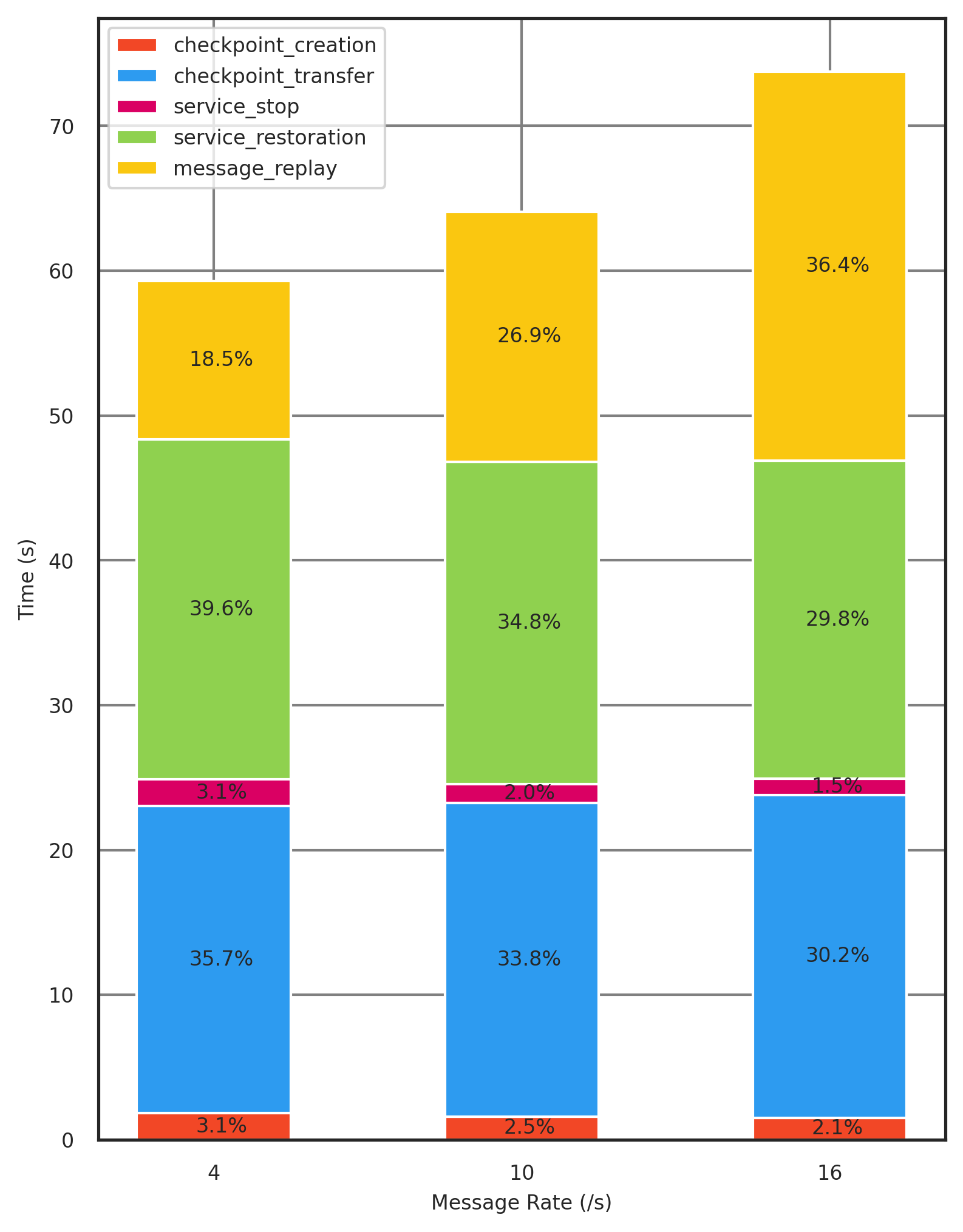}
    \caption{MS2M for StatefulSet Pods}
    \label{fig: Time Distribution Across Migration Phases - MS2M for StatefulSet Pods}
\end{figure}

To better understand the factors contributing to the total migration time for each strategy, we conducted an additional evaluation to examine the distribution of latency across the sub-processes under three different message rates. The results show that as the message rate increases from 4 to 16 messages per second, the proportion of time spent on message replay grows significantly across all strategies, especially for individual pods without the cutoff mechanism. At the highest message rate, message replay accounts for over 80\% of the total migration time in this strategy, making it the dominant factor. The benefits of the cutoff mechanism are clearly demonstrated in Figures \ref{fig: Time Distribution Across Migration Phases - MS2M for Individual Pods} and \ref{fig: Time Distribution Across Migration Phases - MS2M for Individual Pods with Early Stopping}, where the mechanism reduces the time spent on message replay from 80.3\% to 56.2\%, effectively balancing the overall migration duration despite the additional time required for the cutoff mechanism. In the StatefulSet migration, illustrated in Figure \ref{fig: Time Distribution Across Migration Phases - MS2M for StatefulSet Pods}, the service restoration sub-process consistently occupies a large portion of the total time, while message replay gradually increases with higher message rates, reaching 36.4\% at 16 messages per second. Nevertheless, the total migration time for StatefulSet remains significantly shorter compared to individual pods, indicating different dynamics in state synchronization for StatefulSets.

\section{Conclusion and Future Work}

In this paper, we presented our work on the previously proposed MS2M framework, addressing key challenges associated with stateful service migration specifically in modern orchestrators. Our approach integrates Kubernetes' FCC feature to extend support for migrating entire Pods, including those managed by StatefulSets. Additionally, we introduced a Threshold-Based Cutoff Mechanism to effectively handle high incoming message rates during migration, significantly optimizing the process by reducing migration time. The cutoff threshold can be further adjusted based on the requirements for state synchronization time, allowing for finer control over the balance between migration time and downtime.

We evaluated the performance of three new migration strategies—MS2M for individual Pods, MS2M with a cutoff mechanism, and MS2M for StatefulSet Pods—against the baseline Stop-and-Copy mechanism. The results confirmed that MS2M for individual Pods significantly reduced downtime, consistent with previous findings in \cite{9766576}. The previously identified issue in the original MS2M paper—long migration times under high incoming message rates—was addressed by the proposed Threshold-Based Cutoff Mechanism, using a formula to calculate the cutoff threshold $T_{\text{cutoff}}$) based on the maximum acceptable message replay time ($T_{\text{replay\_max}}$). This adjustable threshold enables a balance between downtime and total migration time, enhancing the flexibility and scalability of the original approach.

The integration of MS2M for StatefulSet Pods provided greater flexibility in managing stateful services within Kubernetes, particularly at low and moderate incoming message rates. However, as message rates increase and reach the system’s maximum processing capacity, the benefits of MS2M diminish. This suggests that as the system approaches saturation, the overhead from message replay and state synchronization becomes more significant, limiting the effectiveness of MS2M. Therefore, in high-traffic environments, additional strategies such as scaling out services or implementing load balancing across multiple instances may be necessary to maintain optimal performance during migration.

The results suggest several future research directions. MS2M's limited effectiveness for StatefulSet Pods is partly due to Kubernetes' management of Pods with unique identities. Therefore, developing a new type of controller tailored for stateful microservices could further improve migration efficiency. Advanced algorithms for dynamic message routing and load balancing, particularly in fluctuating network conditions, are another potential direction. Machine learning could also optimize the cutoff threshold, enabling smarter decisions based on real-time metrics like message rates and processing times. Finally, expanding the framework to support other orchestration platforms beyond Kubernetes would broaden its applicability, making it a more versatile solution for stateful service migration in various cloud environments.

\bibliographystyle{IEEEtran}
\bibliography{references}

\end{document}